# THE PROPERTIES AND THE STRUCTURE OF COLLOID AGREGATES OF THE SILVER BESILCATE $Ag_6Si_2O_7$ IN WATER MEDIUM

Germanov N.A. Goliandin S.N. Mochalov S.V. Pulnev S.A.
*"Nanometall" limited, Saint Petersburg, Russia*
*Kompan M.E. Nevedomski V.N. Ulin V.P., Ulin N.V.*
*A.F.Ioffe Institute, Saint Petersburg, Russia*

The paper describes the investigation of the properties of silver besilicate salt colloids in water medium.

In contrast to earlier known hydrothermal method of obtaining crystals of $Ag_6Si_2O_7$, the object of the present study was obtained in a soft conditions, in water medium, at temperatures and pressure close to room ones. The morphology and crystallinity of matter under investigations appear to depend on the growth conditions. Slow growth resulted in quasi-crystalline yellowish whiskers, while high concentration of precursors led to fast formation of the quasi-amorphous aggregates.

The elemental constitution of object under study was determined with the help of Comebax-INCA set-up, and was then proved by chemical analysis. The ratio of elements was close to $Ag_6Si_2O_7$, and depended a little whether the sample was taken from the sediments of reaction or from dried residue of a liquid medium.

The experiments in Raman technique revile several well-resolvable bands in whisker-like samples, but only few of them were able to detect in quasi-amorphous material. In the last case the bands were broader, but the positions of weak broad bands were close to the positions of corresponding bands in whiskers. We can suggest, that the bands, common in two former cases, are due to oscillation in bepyramidal molecules of $Ag_6Si_2O_7$, and the bands, seen only at whiskers, are due to long-rang order.

The very special peculiarity was observed under chemical test with hydrochloric acid. In case of any silver salt one should anticipate the typical reaction – the formation of white snow-like sedimentation of AgCl. It was not the case. But the white flakes sedimentation appeared after addition of NaCl. This result can be explained if to assume the existence of silicon acid shell around the colloidal particles of silver besilicate. The model of silicon acid shell around colloids was proved by high resolution electron microscopy.

# СВОЙСТВА И СТРОЕНИЕ КОЛЛОИДНЫХ АГРЕГАТОВ БИСИЛИКАТА СЕРЕБРА $Ag_6Si_2O_7$ В ВОДНОЙ СРЕДЕ.

Германов Н.А. Голяндин С.Н. Мочалов С.В. Пульнев С.А.
*ООО «Нанометалл» Санкт Петербург, Россия*
Компан М.Е. Неведомский В.Н. Улин В.П. Улин Н.В.
*ФТИ им. А.Ф.Иоффе, Санкт Петербург, Россия*

Серебро – один из металлов, освоенных человеком с древнейших времен; оно использовалось и продолжает использоваться в разнообразном качестве – от материала для украшений и денежного эквивалента до компонентов современных электронных устройств. Свойства серебра и его соединений достаточно хорошо изучены, что отчасти объясняется давним опытом человечества по их использованию. В последние десятилетия интерес к серебру вновь активизировался в связи с применением его в медицине [1] и в фундаментальных исследованиях, как к матрице для мономолекулярной диагностики благодаря эффективному усилению интенсивности неупругого рассеяния света [2-4] в области плазменного резонанса [5].

В данной работе исследовались строение и свойства коллоидных растворов бисиликата серебра в водной среде.

Бисиликат серебра стехиометрического состава $\mathbf{Ag_6Si_2O_7}$ известен [6,7] Ранее этот материал в виде монокристаллов был получен в упомянутых работах прямым синтезом из окислов в автоклавах при температурах 200-580$^0$C. Кристаллический характер полученного материала был подтвержден методами рентгеновской дифракции.

Исследованный в данной работе материал был получен в водной среде, при относительно низкой температуре и атмосферном давлении. Морфология и состав получаемого материала несколько отличались при различных условиях опытов. Частично эти различия будут проанализированы далее.

## Морфология твердых форм бисиликата серебра.

Соединение, далее идентифицируемое как $\mathbf{Ag_6Si_2O_7}$ , было получено при длительной тепловой обработке исходных компонент. После фильтрации и отстаивания, выпаривание раствора приводило к выпадению твердой фазы. В случае длительного синтеза с недостатком исходного кремний-содержащего компонента, твердая фаза выделялась в виде конгломерата одномерных квазикристаллов (рис.1а,1б). Термин квазикристаллы применен, поскольку, по дифракционным данным, дальний порядок в указанных одномерных образованиях слабо выражен. Исследования структуры будут описаны в отдельном разделе.

В случаях, когда синтез проходил в условиях достаточности компонент, макроскопические формы организовываться не успевали; на субмикронном уровне материал имел коралловидную структуру (рис.2).

**Определение элементного состава и химической формулы полученного продукта.**

Элементный состав твердого оранжевого осадка найден методом рентгеноспектрального микроанализа на установке Comebax-INCA; и кроме того, определялся весовым химическим анализом. Исследуемое вещество предварительно подвергалось многократной промывке деионизированной водой и затем высушивалось при комнатной температуре. Полученное соотношение элементов несколько отличалось для случаев, исследовался ли осадок, полученный при синтезе, или сухой остаток, полученный при выпаривании раствора над первичным осадком. Во втором случае относительное содержание серебра было несколько меньше.

Образец первичных экспериментальных данных и таблица относительного содержания элементов показаны на рис.3 а,б,в.

Указанное в таблице соотношение компонент практически не изменялось при варьировании мольного соотношения серебра и кремния в исходных компонентах реакции от 0,5 до 2 , что говорит о образовании стабильного вещества.

Определенное из опыта соотношение компонентов соответствовало брутто –формуле $Ag_6Si_2O_7$.

Структура молекулы, отвечающей вышеприведенной формуле, показана на рис.4. Молекула исследуемой соли представляет собой два кремниево-кислородных тетраэдра, состыкованные одним общим ионом кислорода, т.е. подобна молекуле бикремниевой кислоты $H_6Si_2O_7$ у которой катионы водорода замещены на катионы серебра.

**Неупругое (рамановское) рассеяние света бисиликатом серебра**

Полученный при реакции твердый осадок, в том числе одномерные нитевидные образования, образовывавшиеся в водной среде, исследовались методом рамановского рассеяния. Эксперименты проводились при возбуждении светом He-Ne лазера с длиной волны 6328,1Å на микрорамановском спектрографе MRS320 HORIBA-JOBIN-YVON.

Зарегистрированные спектры (примеры – рис.5а,б) содержали наборы линий от 95 см$^{-1}$ до 882 см$^{-1}$ . Положения линий указаны на рисунке. Спектры, подобные показанному на рис.5а, удалось зарегистрировать как в водной среде, так и у высушенных одномерных квазикристаллов. Показанные на рис.5а спектры дают представление о качестве и воспроизводимости данных для конкретного образца материала. Для образцов, полученных с разным соотношением исходных компонент и в разных режимах положения линий могли несколько отличаться. Для случая бесформенного осадка зарегистрированные спектры показаны на рис 5б. Из сравнения спектров видно, что часть линий в спектре осадка отсутствует, а некоторые , уширенные по сравнению с линиями, показанными на рис.5а, различимы и приблизительно сохраняют свое положение.

В целом, мы полагаем, что такая неопределенность результатов рамановских экспериментов связана с тем, что макроскопические формы вещества, обусловленные процессом кристаллизации/агломеризации, более чувствительны к условиям эксперимента, чем химический состав материала.

Нами не найдены расчеты колебательных спектров подобного вещества в научной литературе. Однако отдельные предположения по интерпретации полос могут быть сделаны. По нашему мнению, наиболее низкочастотная линия 95,5 см$^{-1}$ безусловно отвечает колебаниям с участием самых тяжелых из ионов молекулы – катиона серебра. Наличие дублетов 434-500 см$^{-1}$ и 828-882 см$^{-1}$ может быть связано с расщеплением мод при взаимодействии колебаний двух составляющих молекулу подобных кремний-кислородных тетраэдров.

Также в целях сравнения были зарегистрированы рамановские спектры карбоната и оксида серебра. Они содержали иные наборы линий, не совпадающие с найденными линиям в спектрах бисиликата серебра.

**Исследования структуры дифракционными методами**

Рентгенодифракционный анализ высушенных осадков, полученных обоими способами, показал, что в структуре исследуемых продуктов дальний порядок не проявляется. Это справедливо, по крайней мере, для объемов материала с линейными масштабами более 5-10 нм. На дифрактограммах выделялась интенсивная широкая полоса с максимумом в районе 34 град (при использовании излучения Cu Kα). По угловому положению эта полоса соответствует наиболее интенсивному рефлексу от $Ag_6Si_2O_7$ в моноклинной модификации [6], что указывает на наличие некоторого упорядочения молекул в полученном материале.

**Электропроводность и оценка степени диссоциации соли бисиликата серебра**

Очень важным для понимания состояния описываемого вещества в водной среде является определение степени диссоциации. Степень диссоциации оценивалась по проводимости раствора. Измерения проводимости выполнялись растворе(*) бисликата серебра с содержанием сухого остатка **m = 150 мг/л**. Для того, чтобы устранить возможные погрешности измерений, связанные с тем, что проводимость измерительных устройств электронная, а проводимость раствора имеет ионную природу, измерения проводились по четырехзондовой схеме в широкой области частот. Примененная техника исходно ориентирована на электрохимические исследования, где носит название «импедансометрия». В измерениях использовался импедансметр Solartron S6800. Результаты частотной зависимости импеданса раствора измерения приведены на рис.6. Для дальнейших расчетов величина проводимости определялась в области частот, где измеряемое значение проводимости не зависело от частоты, т.е. ниже 100 Герц.

Как видно из экспериментального графика, величина сопротивления исследованного образца в частотно-независимой области составляет **R = $1{,}8 \cdot 10^6$ Ом**. Это соответствует удельной проводимости раствора **$\sigma = 1{,}15\ 10^{-4}$ ом$^{-1}$ см$^{-1}$.**

-------------------------

(*) здесь и далее для оценки используется предположение, что весь сухой остаток образован солью, и что соль образует истинный раствор. При использовании такого предположения получаемая оценка степени диссоциации является оценкой нижнего предела этой величины.

Исходя из того, что при диссоциации молекулы в растворе будут образовываться пары подвижных ионов - катион серебра и гидроксил, подвижность которых в водном растворе известна и равна **μ = 5,6 $10^{-4}$ $см^2$ / в сек** , и **18,0 $10^{-4}$ $см^2$ / в сек**, соответственно, и считая также, что анион бикремниевой кислоты и заряженные коллоидные частицы имеют существенно меньшую подвижность, получаем для концентрации ионных пар (n) **, n = 1,3 $10^{17}$ $см^{-3}$** . Концентрацию молекул приближенно (см.*) можно оценить в **N = 4,1 $10^{16}$ $см^{-3}$.** Из проведенных оценок следует, что каждая молекула в растворе замещает на протоны минимум половину из имеющихся у нее катионов. Напомним, что это является нижней оценкой степени диссоциации.

**Особенности взаимодействия коллоидного бисиликата серебра с аналитическими реагентами.**

В ходе исследований был обнаружен весьма необычный экспериментальный факт, заключающийся в отсутствии видимого выделения AgCl при добавлении в раствор концентрированной соляной кислоты (растворы остаются прозрачными). В то же время, введение нейтрального раствора NaCl вызывает быстрое помутнение коллоидных растворов, связанное, очевидно, с образованием взвеси из частиц AgCl. Наблюдаемое в кислой среде отсутствие связывания серебра анионами хлора из взвешенных в коллоидном растворе частиц силикатной соли можно объяснить резким снижением степени ионизации и, соответственно, степени гидратации молекул поликремниевой кислоты, окружающих ядро из конгломератов молекул $Ag_{6-x}H_xSi_2O_7$. Необходимо учесть, что протонирование отрицательно заряженных гидратированных кремнекислотных остатков приведет к исчезновению у них заряда и к «схлопыванию» изначально рыхлой оболочки из поликремниевых кислот вокруг наночастиц $Ag_{6-x}H_xSi_2O_7$. Образующаяся плотная пленка будет препятствовать проникновению крупных гидратированных анионов хлора к молекулам силиката серебра. Добавление же поваренной соли приводит к появлению ионов хлора, но не добавляет в раствор положительно заряженных протонов. Ионы хлора в этом случае могут достигать внутренних областей коллоидов и образовывать белый осадок AgCl.

**Электронно-микроскопические исследования**

Исследования методом просвечивающей электронной микроскопии могут рассматриваться как подтверждение вывода о том, что молекулы (или их конгломераты) сильно сольватированы. На снимке ясно видна зернистая структура коллоидного осадка, причем в центральной части зерен коллоида различима более плотная область (рис.7). Это наблюдение подтверждает предложенную модель строения коллоидов.

**Выводы:**

Совокупность проведенных исследований позволяет надежно идентифицировать продукт, полученный в экспериментах по низкотемпературному синтезу, как бисиликат серебра $\mathbf{Ag_6Si_2O_7}$. В зависимости от условий опыта твердый материал может быть получен в форме аморфного осадка или в форме квазикристаллических одномерных образований. В водной среде полученный материал образует коллоидный раствор, при этом молекулы (возможно - конгломераты молекул) окружены оболочкой поликремниевых кислот.

Предварительные опыты показали, что растворы бисиликата серебра обладают противомикробной и противоопухолевой активностью. Исследования в этом направлении будут продолжены.

**Литература**

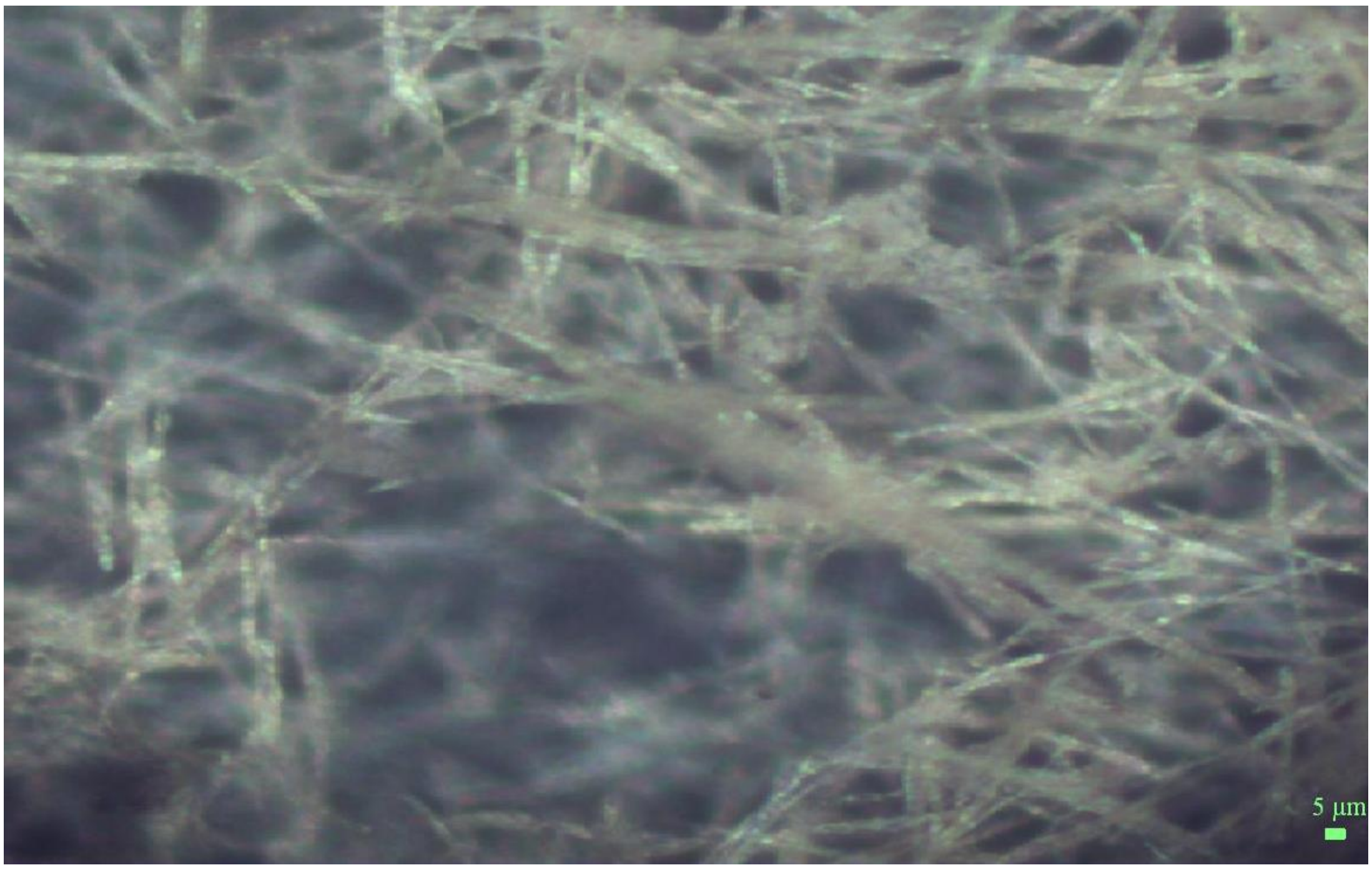

*Рис 1а . Конгломерат одномерных квазикристаллов* $\boldsymbol{Ag_6Si_2O_7}$ *в водном растворе соли. Размерная метка в правом нижнем углу снимка 5 мкм Оптическая микроскопия, микроскоп OLIMPUS BX41, объектив х20 .*

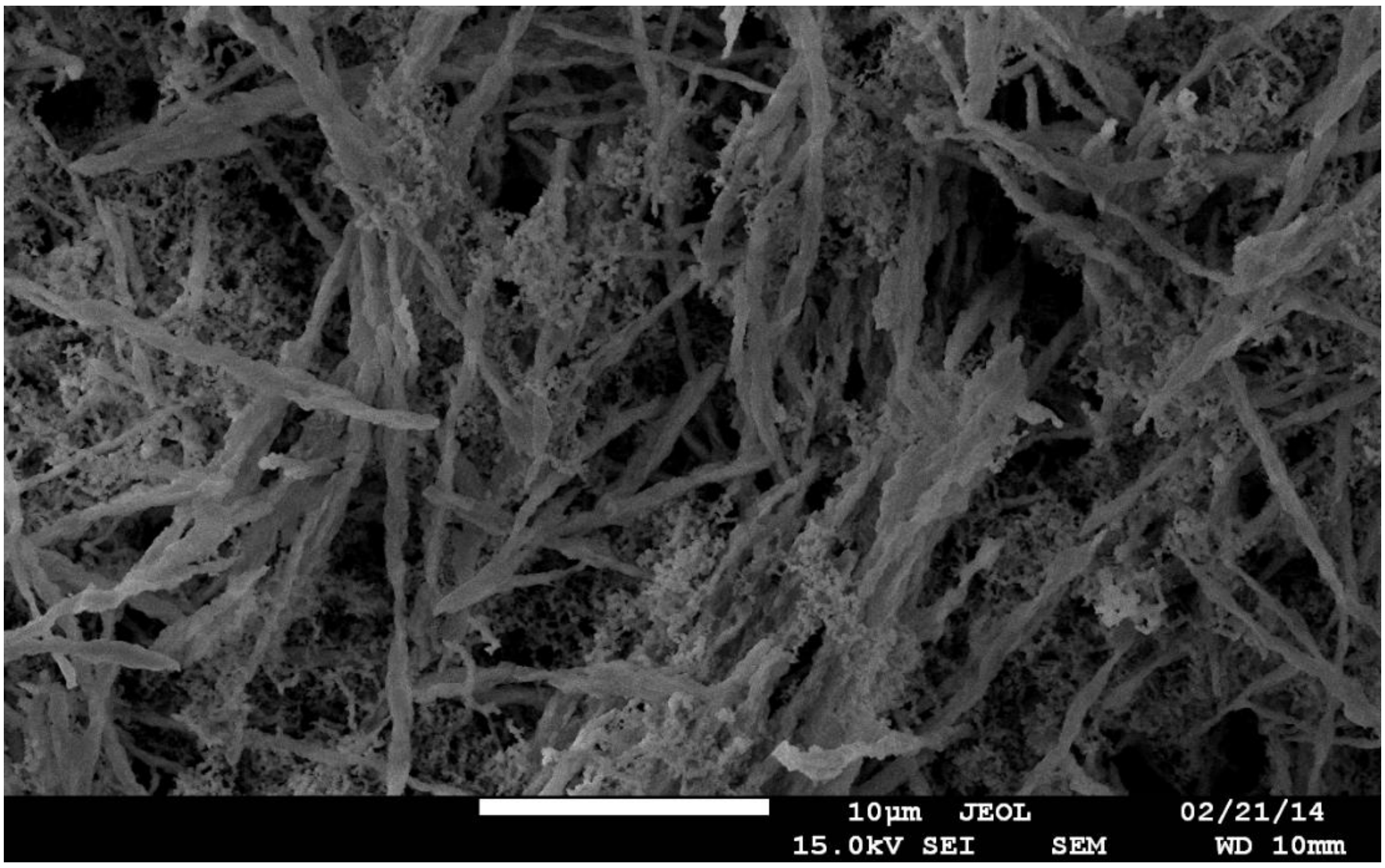

*Рис 1б Электронная микрофотография высушенного сухого остатка из водного раствора соли*

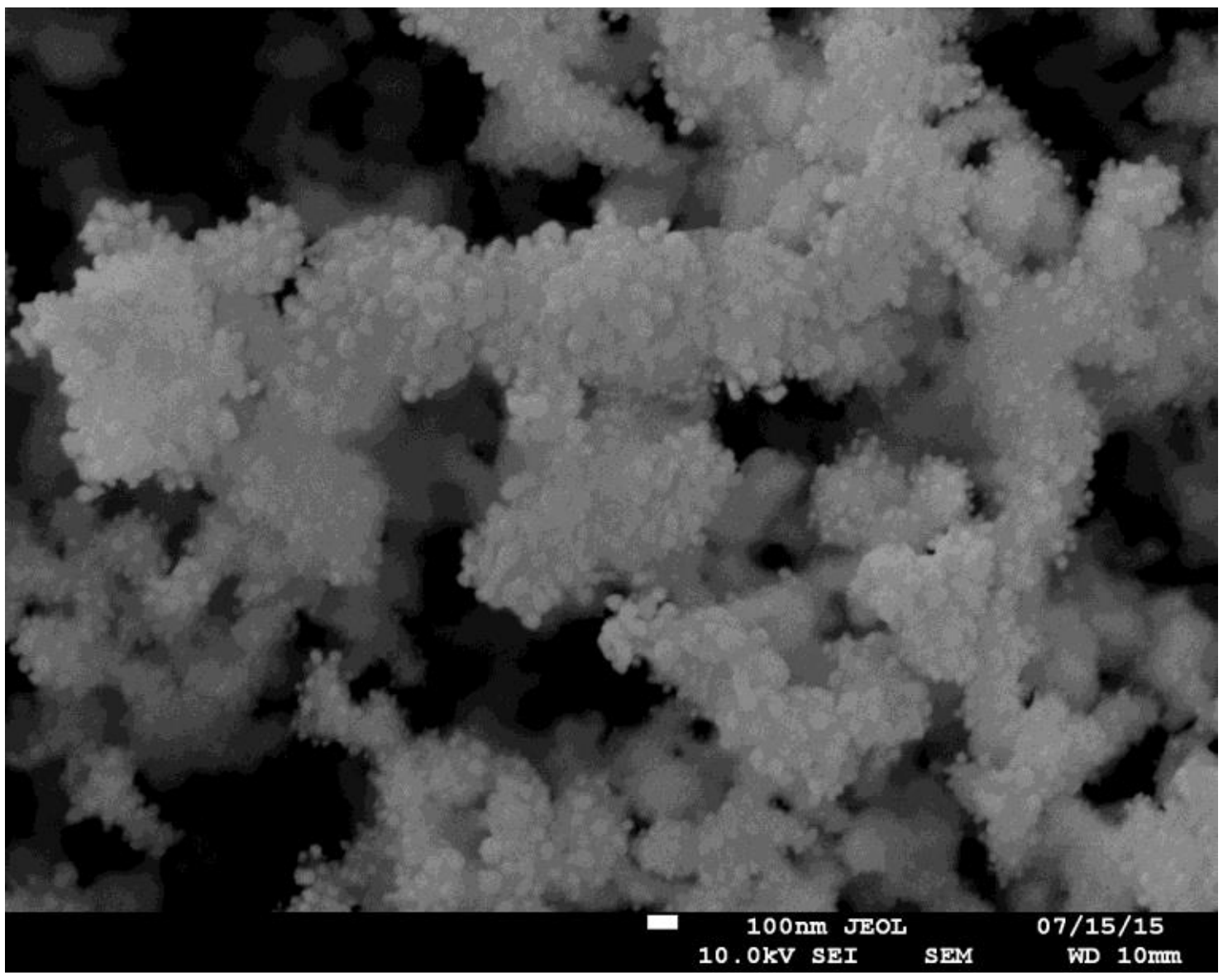

*Рис.2 Морфология твердых форм бисиликата серебра . Электронная микрофотография слабо структурированного осадка*

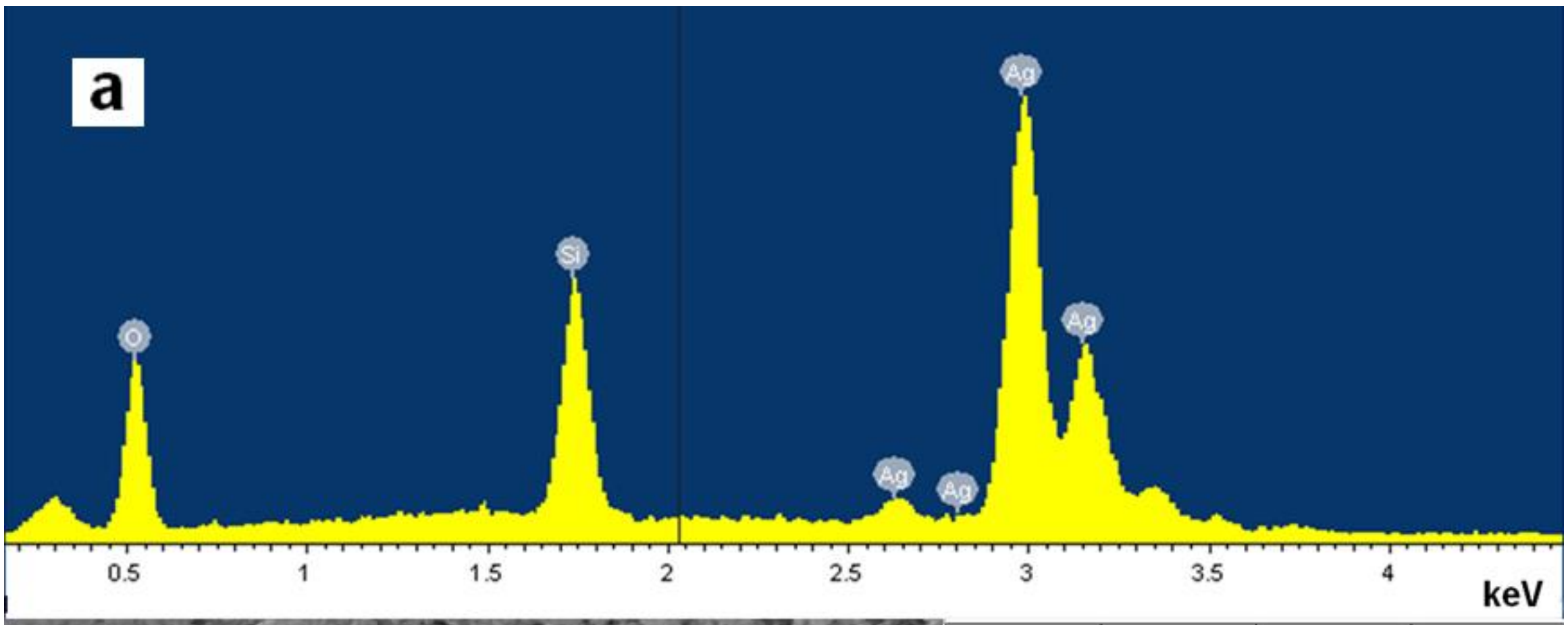

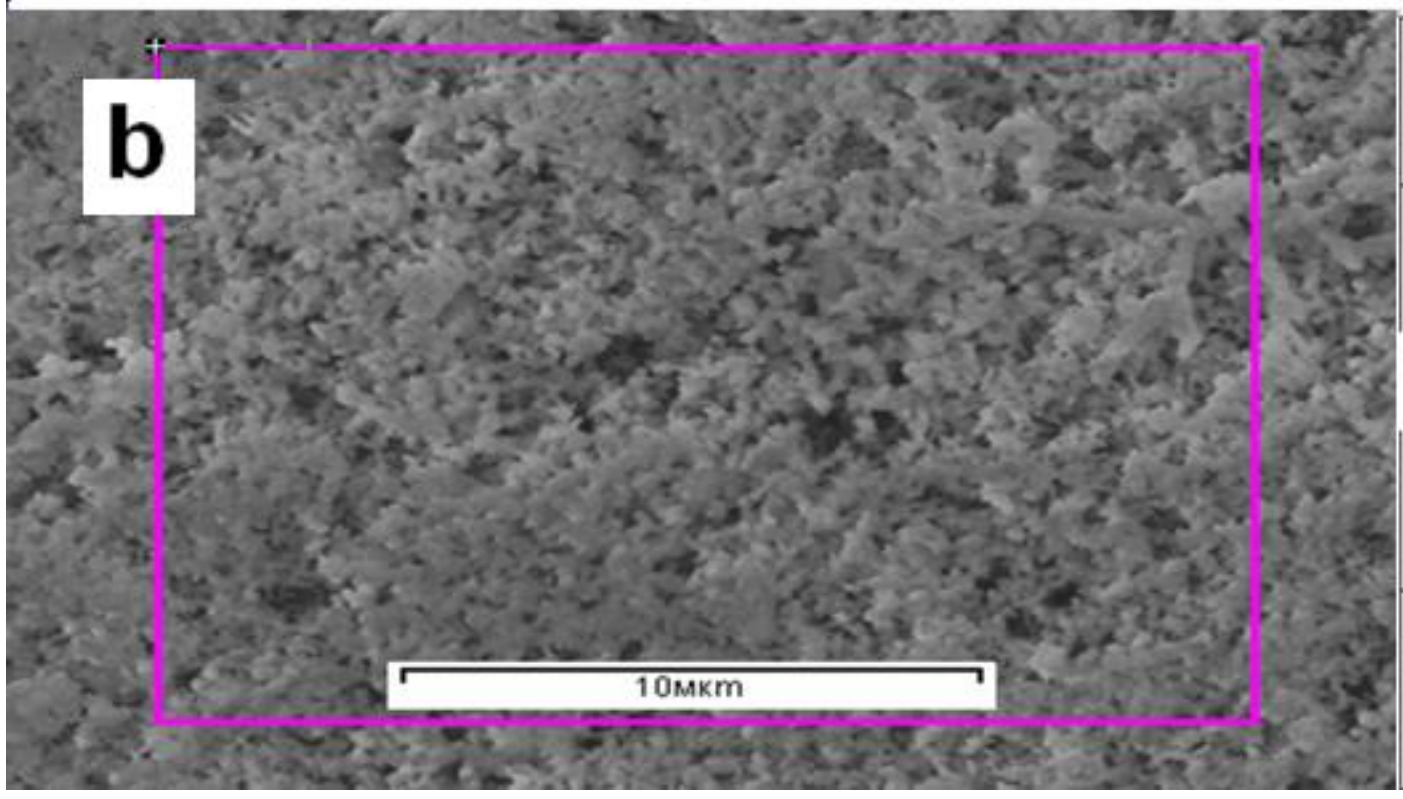

c

| Element | Mass % | Err. mass% | Atomic % |
|---|---|---|---|
| O | 15,31 | 0,43 | 49,39 |
| Si | 7.40 | 0,18 | 13,61 |
| Ag | 77,29 | 0,46 | 36,99 |

*Рис.3 a) - рентгеноэмиссионный спектр исследованного соединения; b) изображение участка материала, спектр которого исследовался; c) таблица относительного содержания элементов.*

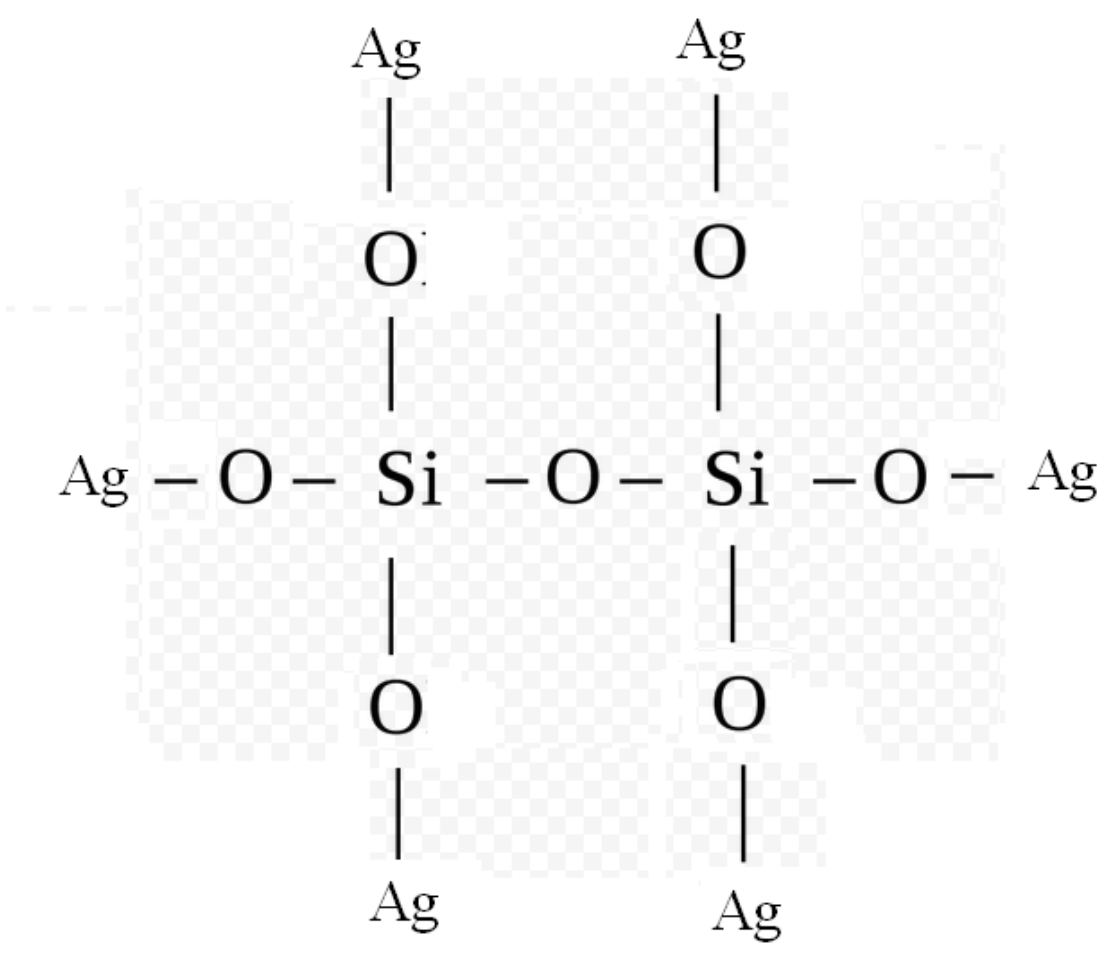

*Рис.4 Структура молекулы бисиликата серебра, соответствующая найденному соотношению элементов.*

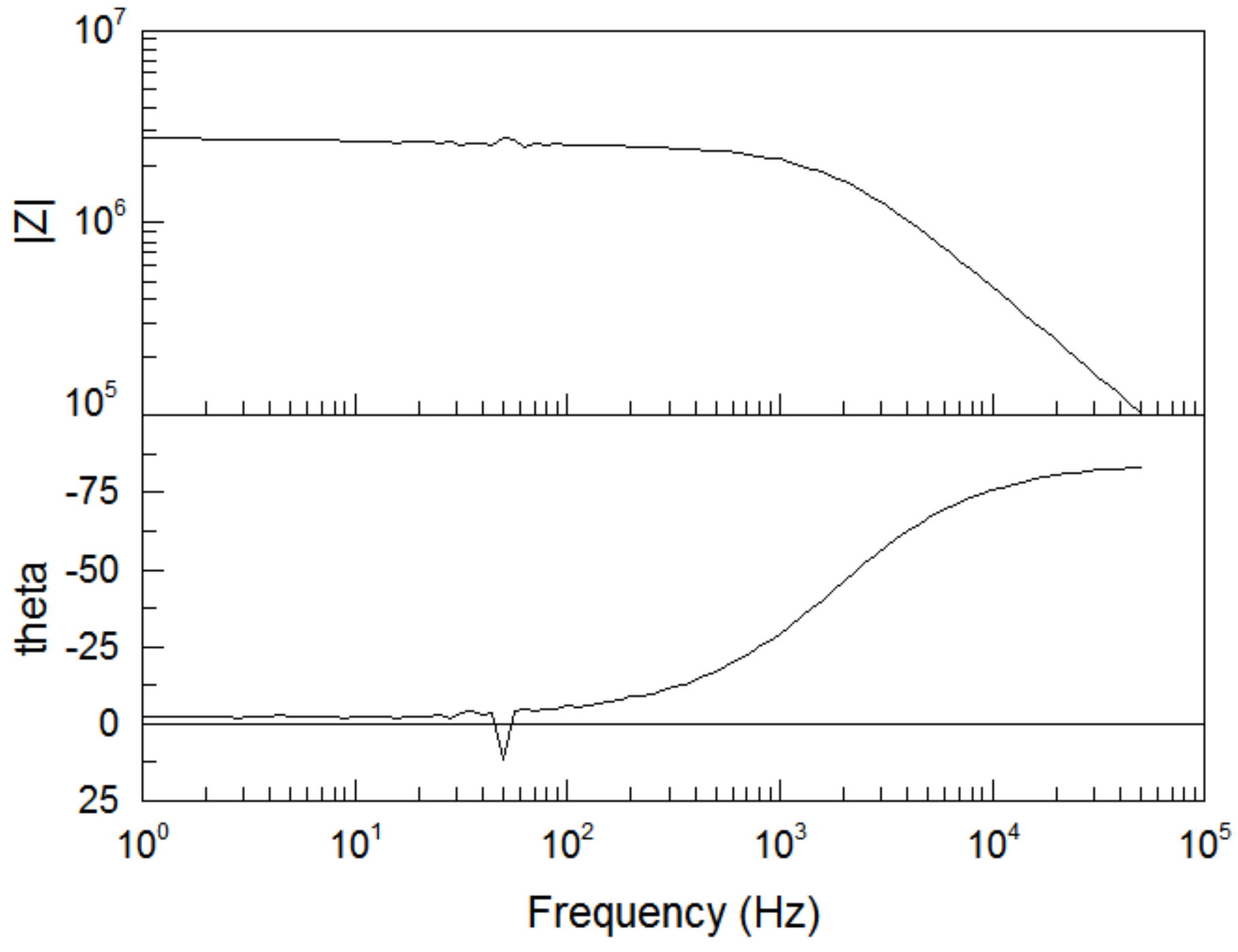

*Рис.6 Частотная зависимость импеданса раствора* $\mathbf{Ag_6Si_2O_7}$ *. Образец – кварцевая трубка с раствором длиной 7,8 см и внутренним диаметром 0,22 см*

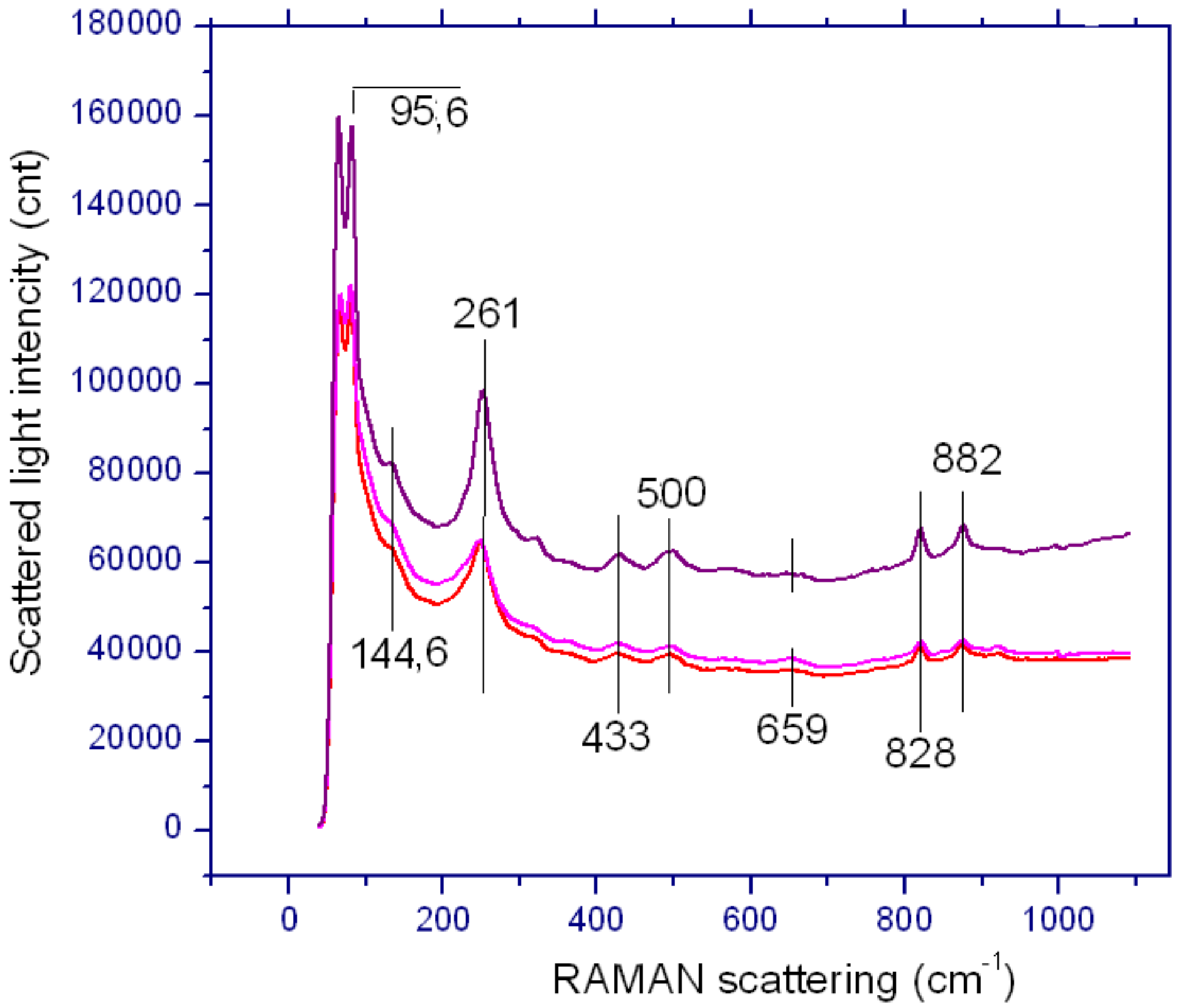

*Рис.5а Рамановские спектры одномерных квазикристаллов бисиликата серебра* $\mathbf{Ag_6Si_2O_7}$

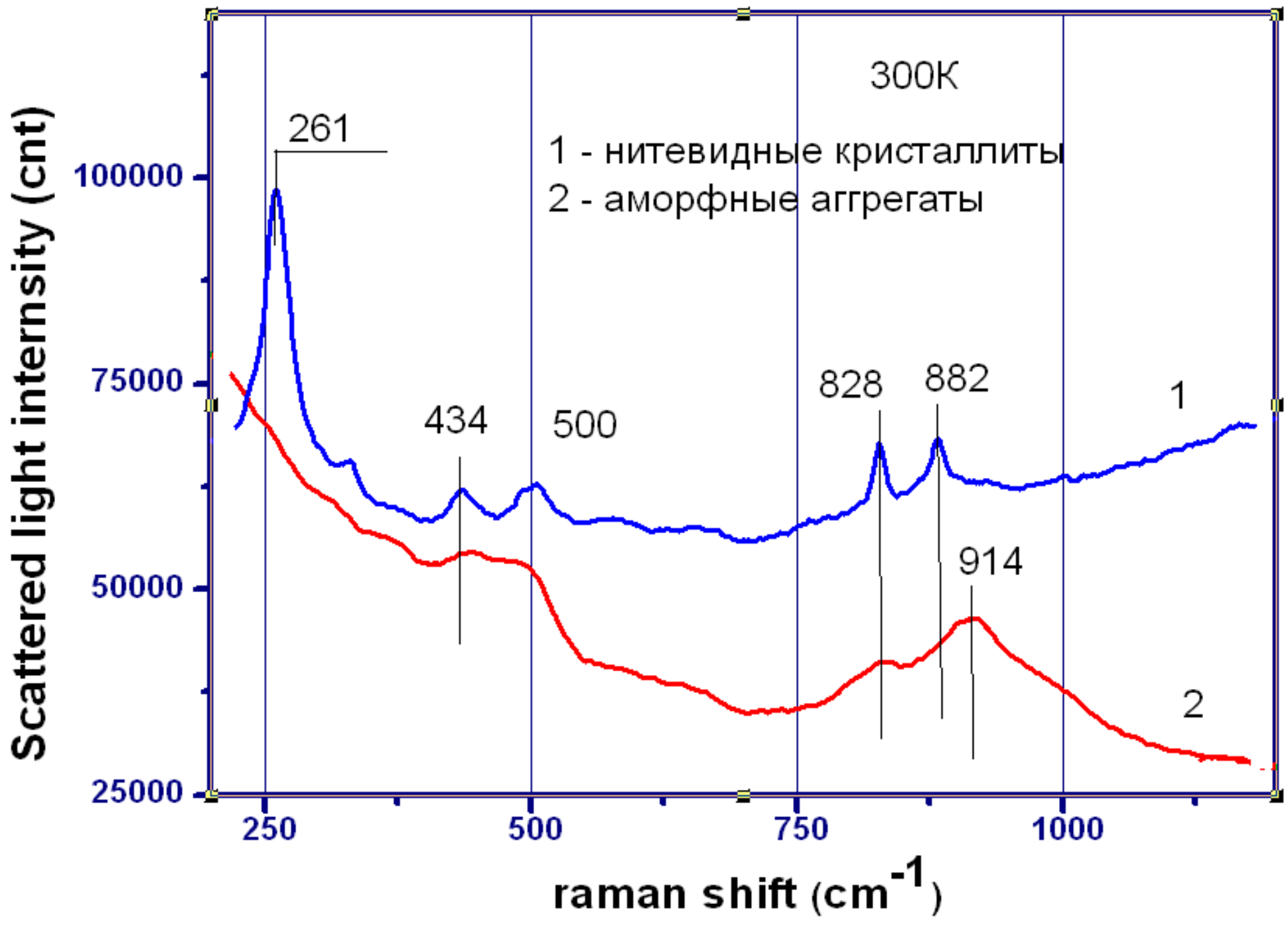

*Рис 5б Сравнение спектров рассеяния света квазикристаллами и аморфным осадком* $\mathbf{Ag_6Si_2O_7}$

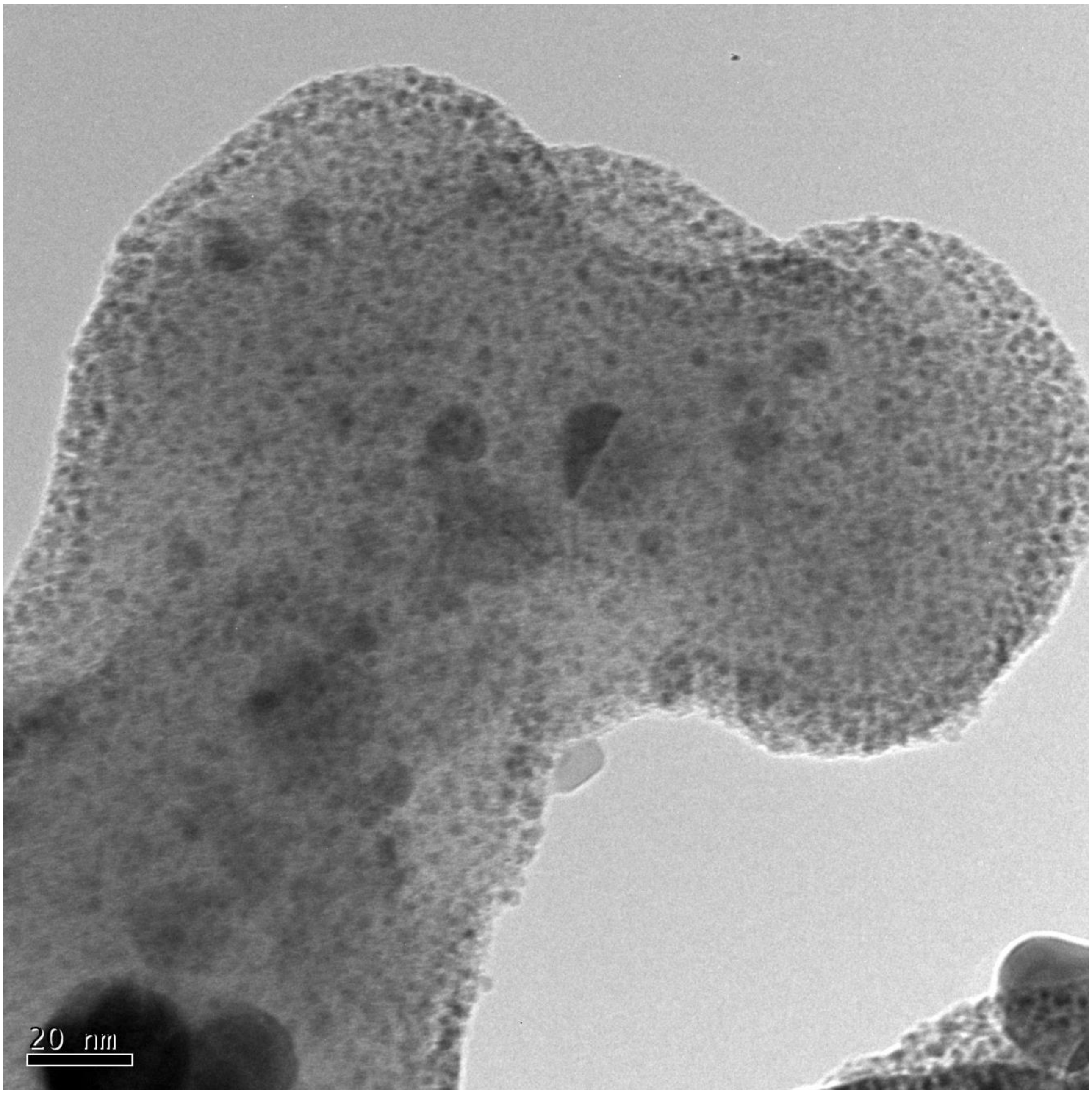

*Рис.7 Микрофотография агрегата коллоидов* $\mathbf{Ag_6Si_2O_7}$